\documentclass[sigconf]{acmart}

\usepackage{balance}

\usepackage{color}
\usepackage{colortbl}

\usepackage{enumitem}       
\usepackage{graphicx}
\usepackage{hyperref}

\usepackage{listings}

\usepackage{makecell}
\usepackage{multirow}

\usepackage{stfloats}

\usepackage{textcomp}

\usepackage{url}
\usepackage{xspace}
\usepackage{xcolor}
\def\BibTeX{{\rm B\kern-.05em{\sc i\kern-.025em b}\kern-.08em
    T\kern-.1667em\lower.7ex\hbox{E}\kern-.125emX}}

\definecolor{mygrey}{RGB}{245,245,245}

\copyrightyear{2026}
\acmYear{2026}
\acmConference[ICSE-SEET '26]{2026 IEEE/ACM 48th International Conference on Software Engineering}{April 12--18, 2026}{Rio de Janeiro, Brazil}
\acmBooktitle{2026 IEEE/ACM 48th International Conference on Software Engineering (ICSE-SEET '26), April 12--18, 2026, Rio de Janeiro, Brazil}
\acmPrice{}
\acmDOI{10.1145/3786580.3786988}
\acmISBN{979-8-4007-2423-7/2026/04}

\begin{document}

\title{On Fun for Teaching Large Programming Courses}


\author{Walid Maalej}
\email{walid.maalej@uni-hamburg.de}
\orcid{0000-0002-6899-4393}
\affiliation{
  \institution{University of Hamburg}
  \city{Hamburg}
  \country{Germany}
}

\renewcommand{\shortauthors}{Maalej}

\begin{abstract}
Teaching software development basics to hundreds of students in a frontal setting is cost-efficient and thus still common in universities. 
However, in a large lecture hall, students can easily get bored, distracted, and disengaged. 
The frontal setting can also frustrate lecturers since interaction opportunities are limited and hard to scale. 
Fun activities can activate students and, if well designed, can also help remember and reflect on abstract software development concepts. 
We present a novel catalogue of ten physical fun activities, developed over years to reflect on basic programming and software development concepts. 
The catalogue includes the execution of a LA-OLA algorithm as in stadiums, using paper planes to simulate object messages and pointers, and traversing a lecture hall as a tree or a recursive structure. 
We report our experience of using the activities in a large course with 500+ students three years in a row.
We also conducted an interview study with 15 former students of the course and 14 experienced educators from around the globe. 
The results suggest that the fun activities can enable students to stay focused, remember key concepts, and reflect afterwards. 
However, keeping the activities concise and clearly linked to the concepts taught seems to be key to their acceptance and effectiveness.

\end{abstract}

\ccsdesc[500]{Social and professional topics~Software engineering education}

\keywords{Fun SE, Programming Education, Student Activation, CS1}

\maketitle

\section{Introduction}

Large introductory courses at universities are particularly challenging for both students and professors. 
In a setting that includes hundreds of students and that is much larger than a workshop or a lab \cite{Pham:ICSEW:18}, communication between the educator and students becomes difficult, hindering the learning process  \cite{Gleason:CT:1986,Rocca:CE:2010,Sass:TP:1989}. 
Students become anonymous amidst hundreds of peers, limiting direct personal interaction. 
Some students might easily get bored or distracted \cite{Douglas:AER:2012, Leustig:19}, e.g.~by their smartphones, by other students, or by the content itself if they find it too simple or too complex. 
This can also happen in an introductory software development course, e.g., when  
students lose focus on abstract topics like recursion or object references, or when a concept is not explained well. 
It becomes challenging to re-engage them and regain their attention. 
Often, even skilled and well-prepared lecturers dislike basic courses \cite{Gleason:CT:1986}. 
In the frontal, plenary part of the course, where the theory and concepts are introduced to all students, it is difficult to engage them, get their feedback, and decide where to slow down or speed up. 

In recent years, computer-supported methods have become popular for coping with such educational challenges. 
For instance, Audience Response Systems \cite{Kay:CE:2009} such as Mentimeter \cite{Mayhew:RLT:2020} enable students to instantly answer questions using their smartphones. 
A lecturer might, e.g., show a source code example and ask about the value of a variable after executing the code. 
In fact, online and computer-supported learning represented a serious alternative to face-to-face teaching already before the COVID-19 pandemic. 
Thousands of Massive Open Online Courses (MOOCs) and education videos about software development have been made available online over the last decade. 
The advantages are obvious: online courses are flexible and can be watched at any time from anywhere \cite{Zhu:OL:2018}. 
During the COVID-19 pandemic peak, many educators had to record their courses, making it hard to argue against offering the recordings afterwards. 
When universities reopened their facilities and students, universities, and societies strongly argued for in-presence teaching \cite{MyPath:22}, a serious question arises: what would motivate students to attend large face-to-face lectures and what is the added value compared to consuming content at home?

Motivated by this question, we designed a novel catalogue of ten \textbf{physical fun activities} which can be used to explain, visualise, and reflect upon programming concepts in a large classroom with hundreds of university students.   
The underlying idea is to ask students to execute entertaining physical activities which metaphorically represent rather abstract programming concepts: such as executing control structures, creating and reusing objects, or visualising object pointers and data structures. 
Lucardie \cite{Lucardie:PSBS:2014} argued that, while fun is generally well-accepted as an effective pedagogical method in the education of children and seniors, it is rarely discussed in university  education literature, which should be further investigated. 
Tews et al.~\cite{Tews:CT:2015} observed that specific fun activities have a significant positive correlation with cognitive engagement of students.

We report on our experience about using the fun activities in a large software development course with $\sim$500 first-year students during three consecutive years. 
We also report on a qualitative evaluation through semi-structured interviews with former students and educators, where we asked about the opinions, impact, advantages, and disadvantages of the activities. 
Section \ref{sec:catalog} describes the fun activities in the order we introduced them in the course with practical hints on how to execute them. 
Then, in Section \ref{sec:evaluation} we introduce the design and results of our evaluation and discuss potential threats to validity. 
Section \ref{sec:relwork} summarises related work, while Section \ref{sec:discusion} discusses the findings and concludes the paper.

\section{Catalogue of SE Fun Activities}\label{sec:catalog}

\subsection{Executing the LA-OLA Wave  Algorithm}\label{sec:laola}
During the first teaching week, most registered students  will participate, even if attendance is not mandatory. 
They are usually curious about the course syllabus, the lecturer, the rules to pass, and their fellow students. 
This is a good opportunity to set up a \textbf{fun culture} for the course by executing a LA-OLA Wave, one of the most famous fun elements in stadiums (see, e.g., \href{https://youtu.be/-lO64a0eOUQ?si=TWrpu7RWGufq_dbw}{\underline{this video}}\footnote{\url{https://youtu.be/-lO64a0eOUQ?si=TWrpu7RWGufq_dbw}} for an example). 
 
\begin{figure}[]
    \centering
    \includegraphics[width=1\linewidth]{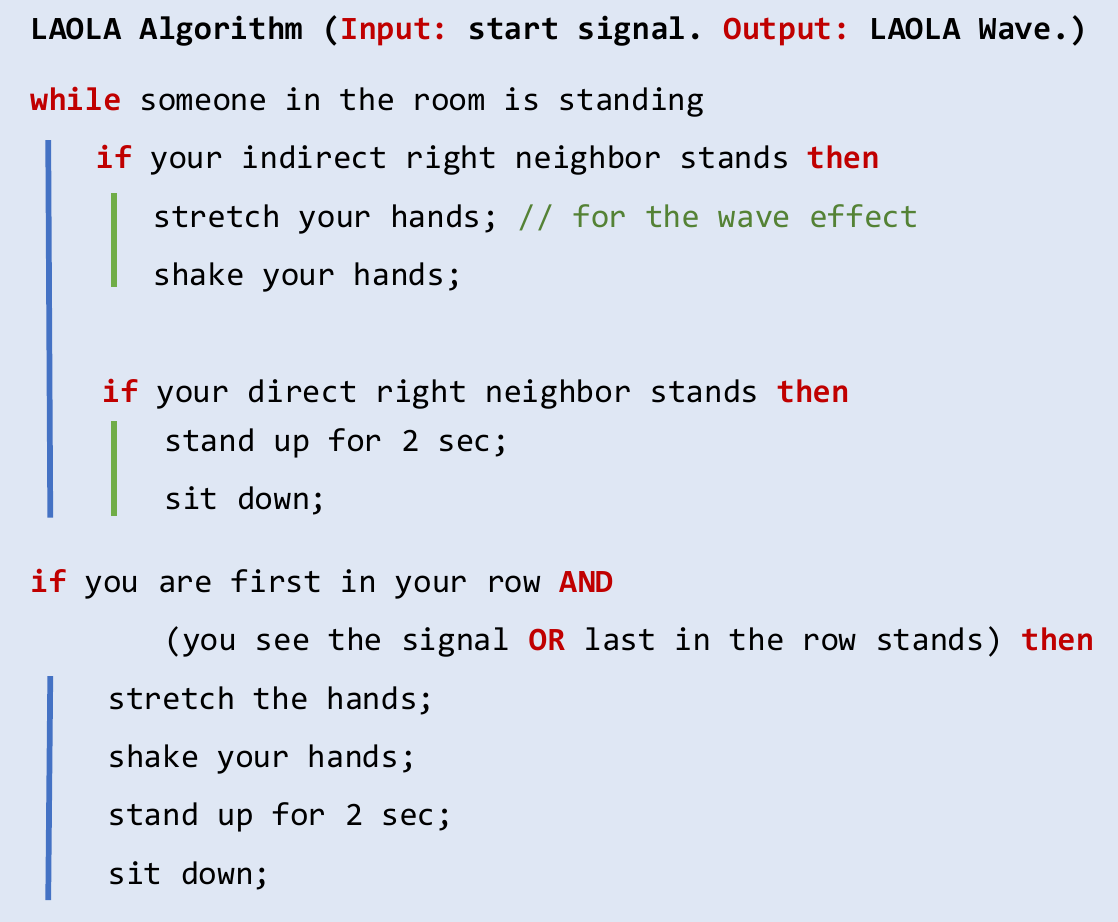}
    \caption{Buggy LA-OLA algorithm to be executed and  debugged with hundreds of students as physical activity, for reflecting on control structures, exceptions, and infinite loops.}
    \label{fig:laola}
\end{figure}

Teaching content of week 1 often includes an introduction into algorithmic thinking and basic control structures, particularly procedure calls, if-statements, and loops. 
For this, LA-OLA represents a suitable fun exercise. 
Figure \ref{fig:laola} shows an example of a LA-OLA algorithm which can be shown to the students on a slide with an example video of how the outcome should look (a short video of a wave in a stadium). 
Then, the algorithm gets executed by the students in the hall. 
The lecturer gives the start signal and observes or records the wave. 
The algorithm shown in Figure \ref{fig:laola} is by design not correct and includes an infinite loop (to introduce this programming concept). 
Reflection questions after executing the wave include: How can we stop the wave/loop (what's the exit condition)? What about exceptions and edge cases, e.g. when some students do not have  neighbours or when they simply miss the wave? 
The lecturer might also introduce the term ``bugs'' in the algorithm. 
A possible homework is to ask students to suggest improvements or alternative implementations and then discuss or even execute those suggestions in later sessions for fun.  

\subsection{Binary Calculator with Student Bodies}\label{sec:calculator}
\begin{figure}
    \centering
    \includegraphics[width=1\linewidth]{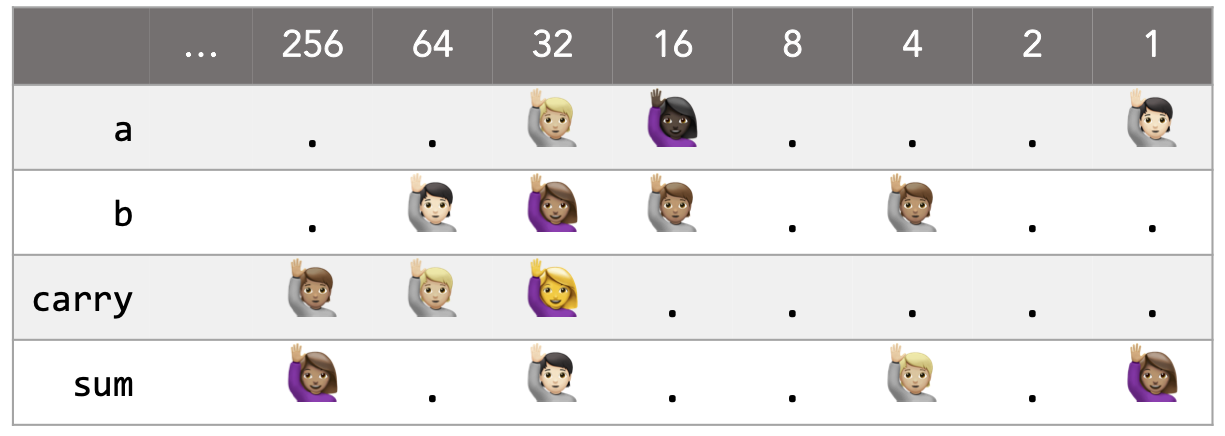}
    \caption{Playing a binary calculator with students. Columns are the bits. A student standing-up represents 1, otherwise 0. The first two rows are the numbers to be added. The third row represents the carry;   the fourth the result.}
    \label{fig:calculator}
\end{figure}

Binary conditions and their combinations represent another basic concept of software development that should be well understood and remembered, and which might be hard particularly for students without prior programming experience or who are not attending a course about logics and theory of computing. 
A short introduction can be given by discussing the binary numeral system (analogously to  the decimal system). 
In a fun activity, students can implement a simple binary calculator as depicted in Figure \ref{fig:calculator}. 
The actual programming assignment should be implemented in the course lab. 
In the lecture hall, the binary calculator can be \textit{physically simulated} by the students themselves. 

Each student represents one bit ($0$ or $1$). 
The first in the row represents the 1st digit in a number, the second the 2nd digit and so on (for instance $a=49$ in the Figure). 
If the bit is $1$, the student should stand up; otherwise, they remain seated. 
The first row in the room should represent the first number $a$ as input to the calculator. 
The second row represents the number $b$ to be added to $a$. 
The third row is the carry, while the fourth row is the sum. 
The fun part emerges from the chaos when students try together to first represents $a$ and $b$, and then going column by column to instruct the carry and the sum to either stand up or to remain seated. 
Depending on the room layout and the number of students, this activity can be designed as a challenge for two or more parallel calculators. 
The calculation algorithm should be explained before the exercise. 

\subsection{Turning the Hall into a Rainstorm with Object-Oriented Software Development}
\label{sec:rainstorm}

\begin{figure}
    \centering    \includegraphics[width=0.6\linewidth]{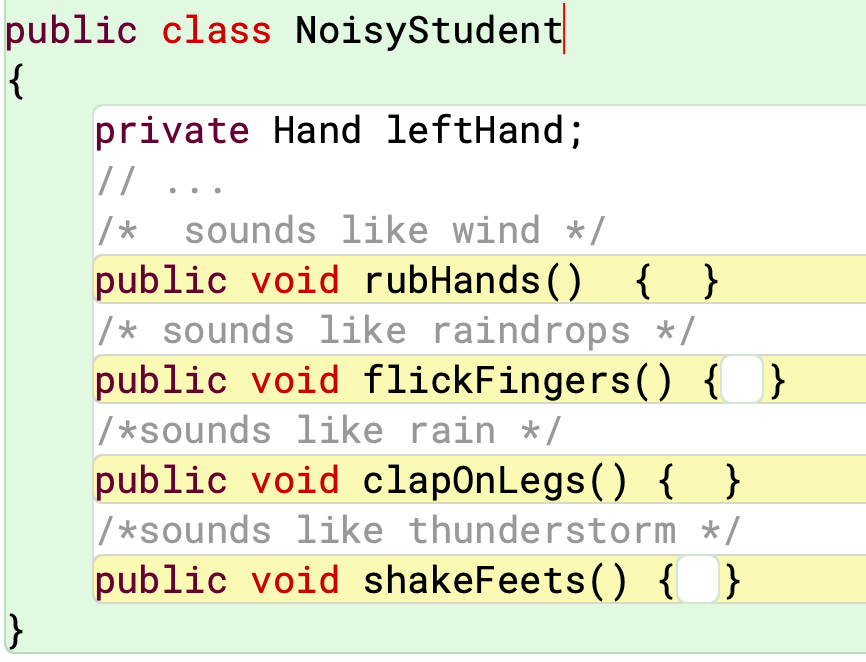}
    \caption{Source code example for collectively creating a sound like a rainstorm in the lecture hall.}
    \label{fig:noisystud}
\end{figure}

A fun activity to activate a large group of people is to ask them to make four simple sounds with their hands and feet. 
If well orchestrated, this can be combined into an impressive rainstorm (see e.g.~\href{https://www.youtube.com/watch?v=UCRQ4qO2rks}{\underline{this video}}\footnote{\url{https://www.youtube.com/watch?v=UCRQ4qO2rks}} or  \href{https://www.youtube.com/watch?v=iT6ttO7yxaE}{\underline{this video}}\footnote{https://www.youtube.com/watch?v=iT6ttO7yxaE}). 
This activity can be used as a metaphor to explain the fundamental concept of an object in Object Oriented (OO) Software Development. 
Objects have states and behaviours (methods) and  states can be changed through the interfaces (method calls). 
In this activity, the object ``a noisy student'' provides four interfaces as shown on Figure \ref{fig:noisystud}: rub hands (sounds like wind), flick fingers (sounds like raindrops), clap on leg (sounds like rain shower), shake feet (sounds like thunderstorm). 
These interfaces are called by the client (instructor) to create a rainstorm sound. 
Clients can interact with the objects only through the interfaces as in OO systems.

The activity demonstrates that objects have independent states even if they are of the same type and expose the same behaviour. 
Multiple objects of the same class (NoisyStudent) can exist at once and might have different states (wind\_sound, raindrops\_sound etc.).  
The client calls can change the object state (e.g.~from wind to raindrops). 
For an interesting acoustic effect, the client (instructor) might instruct half of the lecture hall to keep making the wind sound, while the other half should move to making the sound of raindrops. 
The instructor might also ask the students to actually implement this scenario in the programming language used during the course as homework.

\subsection{Passing Object Messages with Paper Planes}\label{sec:paperplanes}

\begin{figure}
    \centering
        \includegraphics[width=0.5\linewidth]{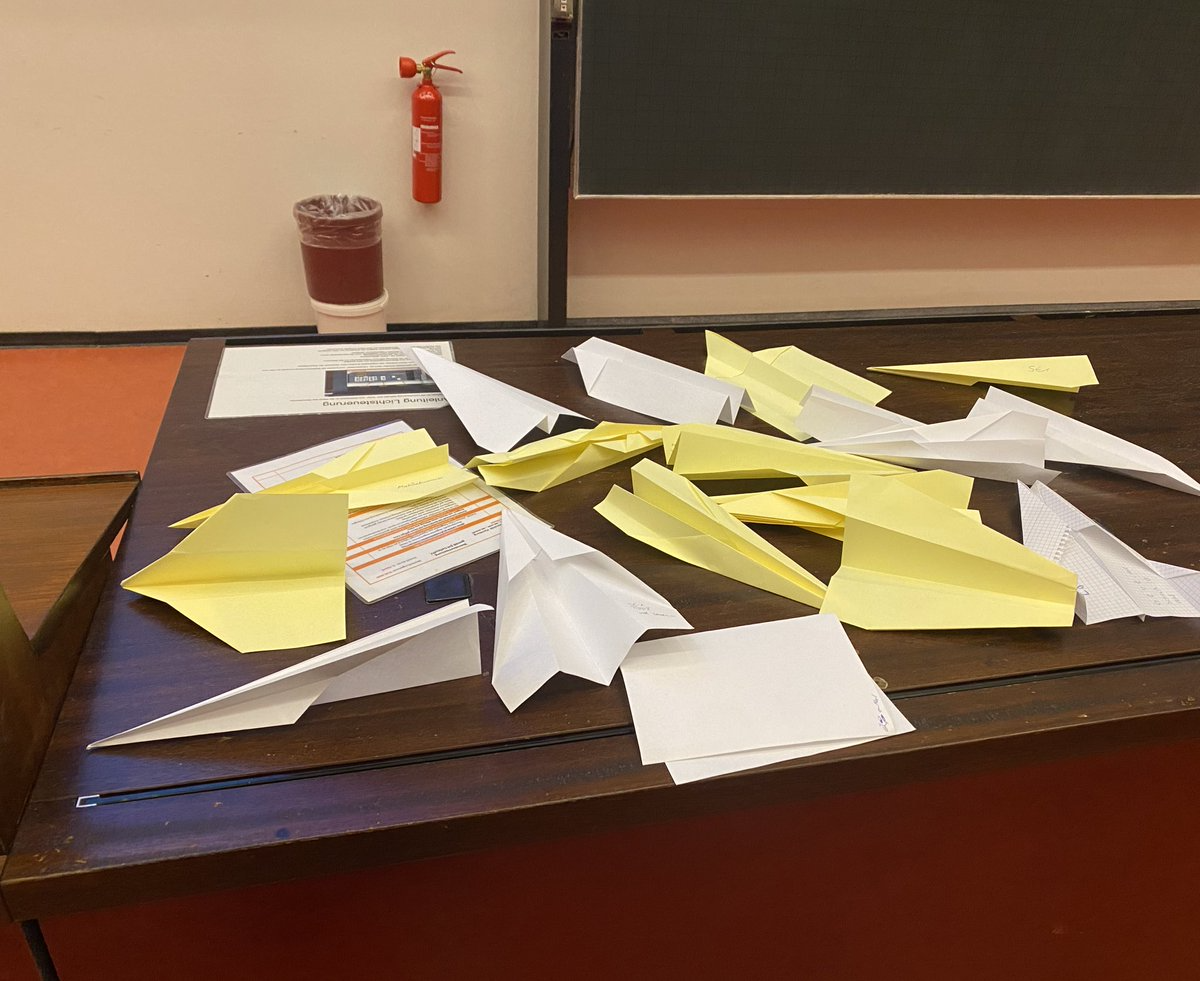}
        \includegraphics[width=0.45\linewidth]{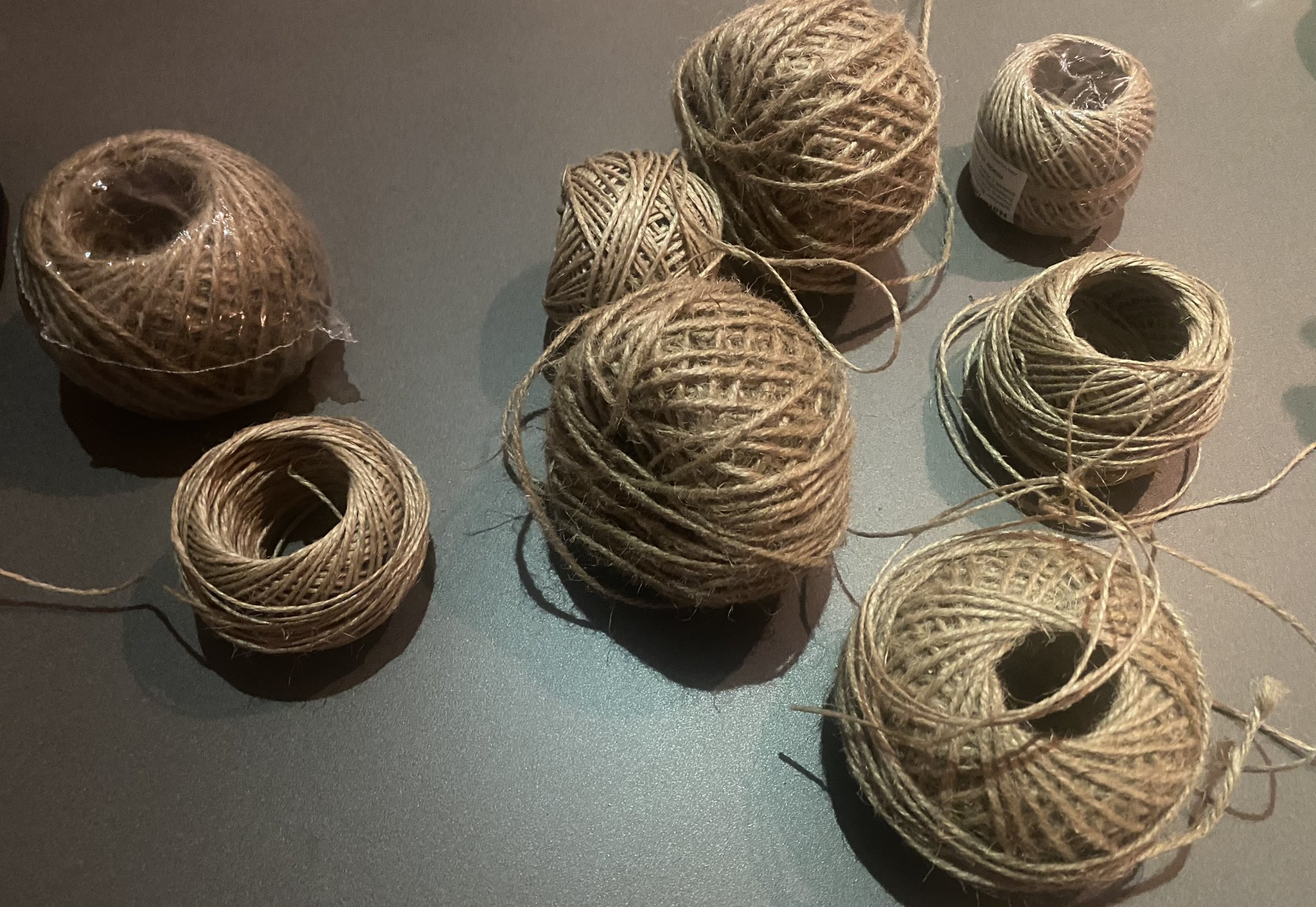}
    \caption{[Left] Paper planes symbolising messages passed between objects (method calls and returns). [Right] Cord symbolising object pointers.}
    \label{fig:paperplane}
\end{figure}

A network of objects and a system snapshot are key abstract concepts in software development that might be hard to grasp for beginners. 
In a network of objects, different objects have references (or pointers) to other objects, through which they can pass messages to these objects or use their services. 
One fun activity is to ask students to call the ``method'' $getStudyTips()$ of fellow students. 
This can be parametrised with the number or the type of tips (e.g., tip related to a certain course, quiet location on campus where to self-study, or which stipends to apply for...). 
Students can write the actual parameter of the call on a paper  plane using a pen. 
Throwing the plane symbolises a method call. 
The return value might itself be a paper plane with a different colour (yellow on Figure \ref{fig:paperplane}). 

However, without a reference (object pointer) between the sender and receiver, the message will get lost in the hall. This becomes a mess and fun at once. 
To visualise pointers, the instructor can use cords (e.g., wool balls as shown on Figure \ref{fig:paperplane}) which the students can throw at a fellow student from whom they want the tips. 
This enables the visualisation of an object network in the room, as system objects can also be placed anywhere in the computer memory and are wired with the references. 
It is also a good opportunity to reflect on the different ways of getting a pointer: by creating the object, through passing the pointer via a parameter, or by saving the pointer in a reference variable). 
Additional concepts such as the alias problem can be discussed with this activity as well. 
A variation is to throw all paper planes to the instructor, where no explicit pointers are needed, symbolising a Singleton and a single point of failure.   

\subsection{Black-Box Reuse with The “NumNum Cat”}\label{sec:numnum}
Animal videos, particularly cat videos, are very popular among students and in social media in general \cite{Maddox:NMS:2021}. 
The NumNum Cat is a popular short  video clip (few seconds) of a tiny cat drinking milk and making a strange ``nuamnuam'' sound. 
Many creative people used this clip as background beat to create impressive soundtracks (such as \href{https://www.youtube.com/watch?v=GArzu9ttQ0M}{\underline{this video}}\footnote{\url{https://www.youtube.com/watch?v=GArzu9ttQ0M}} 
or \href{https://www.youtube.com/watch?v=_C1FZ4HtzGY}{\underline{this video}}\footnote{\url{https://www.youtube.com/watch?v=_C1FZ4HtzGY}}).
This can be used as a funny, unexpected and memorable example of black-box reuse (or API usage), an important fundamental concept in software development. 

The video example enables reflecting about multiple aspects. First, creativity is important in software reuse as the solution is often not trivial and  APIs need to be ``repurposed'' or ``wrapped''. 
Second, black-box reuse means that developers must live with the available functionality. 
In this case, the NumNum clip can be integrated into the soundtrack, but not changed---similar to reusing a software library. 
Third, APIs can be misused \cite{Zhang:ICSE:2018} (or used for not intended purposes). 
As a  complementary homework challenge, the lecturer can share a simple Java class of a cat that can play several simple sounds on the computer including the NumNum sound. Students can then reuse this class, without changing it, to create complex sound tracks using some complex iterations. 
The cat might also paint simple pixels on the screen and the students can be asked to create creative complex visualisations by reusing the cat interfaces.

\subsection{Recursive Counting with Students}\label{sec:recursion}
Recursion is among the most difficult concepts for students to understand \cite{Haberman:ITCSE:02,Lacave:TE:17}. 
There are plenty of good examples that can be used to explain recursion, particularly when the problem itself is recursively defined (e.g.~calculating Fibonacci numbers). 
An alternative fun activity is to ask students to solve a physical task in the classroom recursively. 
For instance, a random number of students might line up, e.g. to get a small gift from the instructor (e.g. a pen or a chocolate bar). 
Now, since the number of gifts is limited, the last student entering the queue would like to know how many people are lining up without walking to the front and doing the entire count. 
The recursive implementation would be to recursively ``call'' the students in the line to count until they arrive at the first student in the line. 
This student represents the base case (termination) and should return zero as no one is in front. 
The next will return +1; the next +1, etc.

\begin{figure}
    \centering
    \includegraphics[width=1\linewidth]{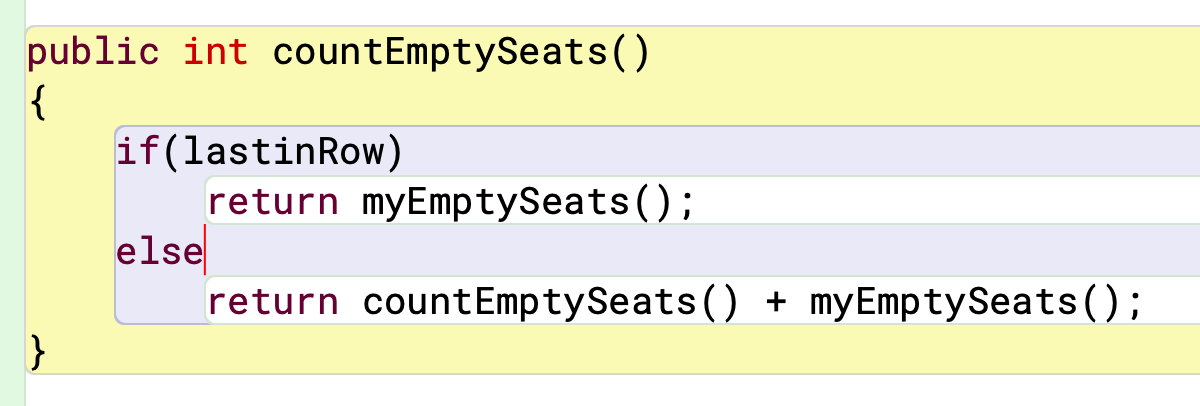}
    \caption{Students recursively counting empty seats in a row.}
    \label{fig:emptyseats}
\end{figure}

A more complex and fun task is to ask students to recursively count empty seats in the rows where they as sitting without standing, as shown in Figure \ref{fig:emptyseats}. 
The students will call $countEmptySeats()$ recursively. 
The last one (sitting next to the corridor) will return the number of empty seats until the corridor; the previous will return the sum of what the last student returned and the empty seats between both and so on.
This example enables reflecting on the Divide and Conquer strategy as the problem becomes smaller and thus easier to solve with each step. 
It also nicely visualises (with the students) the call stack of the callbacks. 
One way to vary this activity is to ask several rows in the hall to do the count as a  competition and to observe potential mistakes, which will certainly happen if students don’t carefully execute the algorithm.

\subsection{Linked Lists with Finger Pointing}\label{sec:linkedList}
The last part of our course introduces basic algorithms and data structures frequently used in software development. 
This is particularly important when students  come from different disciplines (including CS minors) and some of them do not have a dedicated course about algorithms and data structures. 
In addition to a quick introduction into the concepts of lists, sets, queues, stacks, and hashmaps, the students learn to use common APIs (in our case the Java Collection Framework) as well as understand their basic implementations.
One of those are array lists and linked lists, including simply and doubly linked with their advantages and disadvantages. 

While the concept of arrays is straight and can be mapped to physical examples in every-day's life (such as storage warehouse, a row of seats, or a numbered bookshelf), the concept of linked lists might be abstract for beginners. 
A fun activity for explaining linked lists is to create one by asking volunteer students randomly seated in the hall to simulate the $ListElement$ objects by pointing to (and noticing) their successor elements in the list (and predecessor for double-linked list). 
Once the list is constructed, it can be traversed, e.g.~to find an element. 
Elements are added and removed from the list, which physically shows how fast these operations can be conducted and what exactly happens: i.e.~which pointers are added, removed, or changed. 
This also nicely shows the efficiency of the operations compared to an array where all successors need to be shifted if an element is, e.g., deleted.  

\subsection{The Tree in the Lecture Hall} \label{sec:traversing}
Trees and graphs are other important data structures that software development students should learn and understand well. 
The layout of the lecture hall can be used to represent a tree or a graph in a fun activity. 
The lecturer can play the root node of a tree and students in the follow-up  rows can play the children nodes. 
Each student should first establish a reference to its parent node.
The educator then develops with the students a strategy to \textbf{traverse} the whole hall structure. 
Visiting a node in the tree (i.e.~a student) corresponds to a clap. 
This exercise can be used to discuss the advantages and disadvantages of the different data structures. 
If a tree is used, different root nodes might be tried out to finally identify a more balanced tree (and a faster traversal). 
This gets nicely reflected in the sound of how many people are clapping at once.
Disadvantages of unbalanced trees can be discovered and discussed, and strategies of re-balancing the trees can be introduced.

\subsection{Quick- and Bubble-Sorting Students} \label{sec:sortingstudents}
Visualisation is key to understanding and remembering complex algorithms and underlying key ideas such as sorting algorithms \cite{Shaffer:CSE:07}. 
One fun activity to visualise sorting algorithms from different complexity classes is to ask two teams of volunteers to execute in the hall two different sorting algorithms, e.g.~bubble sort and quicksort. 
The sorting criterion can, e.g., be the height of students, which can easily be compared between two students. 
Another criterion might be the birth month of the students (e.g.~enabling to discuss stable vs.~unstable algorithms if two students are born in the same month). 
The goal is to get a row of students sorted by strictly following the algorithm.
This activity helps understand the algorithms and reflect about their different computational complexity.
An algorithm should be introduced to the students before asking them to execute it. 
Stopping the time will show the difference between the algorithms even if students try to act faster.

\subsection{SOS: The Symphony Orchestra System}
\label{sec:symphony}
This fun activity represents a highlight and can thus be conducted during the last session.
The goal is to build a ``Symphony Orchestra System'' with the help of the students using simple means and then reflect about the system dynamics. 
The core idea is to ask the audience to play simple rhythms (or music tones) using Boomwhackers or similar devices. 
Boomwhackers are coloured plastic tubes producing different musical notes (see e.g.~\href{https://www.youtube.com/watch?v=e5n1-xDT6R8}{\underline{this video}}\footnote{\url{https://www.youtube.com/watch?v=e5n1-xDT6R8}}). 
By distributing different colours to different student blocks in the hall, the room can be repurposed to a large ``piano''. 
For a less expensive alternative, the lecturer might create such devices from cardboard tubes by cutting them in different lengths, which lead to different sounds when clapped on the hand. 
The easiest option (with lower quality sounds) is to ask students to use similar objects they carry with them to create different noises (e.g.~pens, plastic bottles, metal bottles, etc.). 

This activity can be used to reflect about different software development topics, such as algorithms (for getting complex rhythms or sounds), loops (for refrains), object states and behaviours (for getting simple sounds and rhythms), controller objects to orchestrate the music, agile development (improving the sound and increasing complexity in each iteration), or even multi-threading and parallel execution if this is part of the course. 
Also, the concept of inheritance can be explained with this activity, since the different tubes inherit the characteristics of an (abstract) tube but have different specific sounds. 
Recently, Zhang et al.~\cite{Zhang:ITCSE:2022} discussed how Coded Music Composition can be used to teach computational thinking. This can inspire sound compositions in this activity.  

\section{Course Evaluation and Interview Study}\label{sec:evaluation}
Evaluating the effect of fun activities on teaching is non-trivial since fun does not represent a goal by itself and is usually a ``vehicle'' for higher motivation and participation. 
Particularly, \textit{isolating} and quantifying the effect of our activity catalogue on learning success and efficiency would require controlling the course environment over multiple cycles (some cycles with and other without the activities) as well as systematically collecting student demographics, success, and engagement data to control for confounding factors. 
Both prerequisites are unfeasible in our setting due to strict data collection policies as well as high fluctuation of the students' backgrounds and study majors. 
Moreover, the course content, the lecturers' experience, and teaching skills are continuously evolving, which hinders comparability across multiple years. 

We first introduced the fun elements after two years of online-only education during the COVID-19 pandemic peak period, which makes comparability between the years even more flawed.  
Moreover, as we assume the activities will be beneficial for the course and the students, refraining from conducting them, e.g. with a subset of the students would be ethically and legally challenging. 
The primary obligation is to teach the course in the best possible and feasible manner and treat all students equally. 
Therefore, we decided to focus on a qualitative evaluation. 
First, we analysed the course evaluation results and compared them to the results from the years before the pandemic. 
Second, we conducted an in-depth interview study with students who participated in the course as well as professors who teach similar courses. 

In our university, the evaluation form changed during the pandemic: it became online with fewer and slightly different questions. 
This makes the direct comparison of quantitative metrics such as satisfaction rates with instructors or self-estimated learning progress  impossible. 
One question was still comparable between the years, though. 
This was about giving the course an overall grade between 1 (excellent) and 6 (very bad). 
Here, we did not observe a significant change between the iteration before the pandemic (without the fun activities) and the three iterations after (with). 
The three median grades were 2, with similar means and standard deviations.

When analysing the free-text questions, where students can highlight the most positive and negative aspects of the course, we made an interesting observation. 
Before introducing the fun activities, the most frequently stated positive keywords were ``nice/kind/helpful atmosphere'' mentioned 19 times, followed by ``practical exercises'' mentioned 17 times (in 72 positive feedback items filled by 97 students). 
After introducing the fun activities, the most frequently mentioned positive feedback was ``interactivity/fun'' mentioned in 7, respectively 14 times followed by ``clarity of content'' mentioned 5 times (in 19, respectively 18 positive feedback items filled online by 46 respectively 62 students). 
In the negative feedback, we did not notice any major difference. 
There were no points directly related to the fun activities. 
In the open question related to improvement suggestions, one student stated that the lecturer should first focus on fully answering the questions of the students in detail and then perhaps conduct a ``game''.

Moreover, we conducted a regular quick survey during the second and third trial iteration every 2-3 weeks asking the following question: ``How do you like/dislike the fun elements during this course?''. 
Students could select one of 5 answer options ``Good'', ``Rather good'', ``Neutral'', ``Rather bad'', ``Bad''. 
Throughout both courses the results were consistent with $\sim$70-80\% of responses being positive or rather positive, 15-22\% neutral, and 5-9\% negative or rather negative (N=120-240 participating students). 
This confirms that the elements were well perceived by the majority during the entire course (and not only at the beginning of the course).

\subsection{Interview Design}

\subsubsection{Interviews with Students}
While the course evaluation gave a first hint about the perception of the students towards the fun activities, it remains rather generic and superficial since neither a focus on the fun activities nor follow-up clarification questions were possible. 
Thus, we decided to conduct in-depth, one-to-one, semi-structured interviews with students to gain more insights about whether and how the activities helped. 

To ensure interviewees are free and unbiased, we recruited students only \textbf{after the course} and all its exams had finished. 
We reached out to 20 students on campus during the follow-up semester and asked if they would be willing to participate in short interview sessions. 
Initially, all of them agreed and shared their email addresses, but only 15 finally participated (10 who attended the first iteration and 5 the second).

Each interview session was conducted via zoom and lasted for about 15-20 minutes. 
After a brief introduction and consent collection, we quickly listed  (1-2 minutes) the fun activities to ensure the students recall them. 
Then, we asked the following questions:
\begin{enumerate}
    \item How did you like/dislike the fun activities and why?
    \item Did they help with learning/passing the course? How?
    \item What were the activities' advantages and drawbacks?
    \item Were the fun elements the main reason for you to attend physically (despite the shared videos)?
\end{enumerate}
We closed the interviews by asking if the students had any other related comments or feedback.
We simultaneously typed the responses in a prepared form and followed up with clarification if feedback was unclear or not detailed enough. 
We briefly summarised our notes at the end of each question to ensure a correct understanding.

\subsubsection{Interviews with Educators}

\begin{table}
    \centering
\small
   \caption{Overview of interviews with educators.}
    \begin{tabular}{p{0.025\linewidth} p{0.115\linewidth} p{0.114\linewidth} p{0.58\linewidth}} \\
\hline
      \textbf{ID}   & \textbf{Region} & \textbf{Univ.} & \textbf{Teaching backgrounds (course size)} \\
\hline
\rowcolor{mygrey}
        P1 & Europe & Research oriented & Large programming \& software engineering courses, no data structures ($\sim$500 students)\\
        P2 & Europe & Teaching oriented & Large object-oriented programming courses (up to 300 students) \\
\rowcolor{mygrey}
        P3 & North America & Research oriented & Large programming courses and software engineering ($\sim$200 students)  \\
        P4 & Europe & Research oriented & Large programming and software engineering courses (up to 400 students) \\
\rowcolor{mygrey}
        P5 & Europe & Research oriented & Large courses on algorithms, data structures and programming  (up to 400 students) \\
        P6 & North America & Research oriented & Large programming and software engineering courses(up to 400 students) \\
\rowcolor{mygrey}
        P7 & Middle East & Teaching oriented & Large programming courses including logic (up to 150 students) \\
        P8 & Asia {Pacific} & Research oriented & Large software development and architecture courses (up to 150 students) \\
\rowcolor{mygrey}
        P9 & Africa & Teaching oriented & Programming, data structure and databases ($\sim$100 students) \\
        P10 & South America & Teaching oriented & Very large courses on C and OO programming (up to 1000 students)  \\
\rowcolor{mygrey}
        P11 & South America & Teaching oriented & Very large courses on C, OO programming and data structures (up to 1000 students) \\
        P12 & South America & Research oriented & Programming and OO software development ($\sim$120 students)  \\
\rowcolor{mygrey}
        P13 & North America & Teaching oriented & Intro into programming, intro into software development (up to 400 students) \\
        P14 & Central America & Teaching oriented & Intro into software development (all sections up to 150 students) \\
\hline
    \end{tabular}
    \label{tab:interviewees-profs}
\end{table}

To also cover the perspective of educators and get external feedback, we interviewed 14 computer science professors, whose backgrounds are shown in Table \ref{tab:interviewees-profs}. 
We focused on recruiting educators who are involved in basic foundational teaching and who had previous experience with programming and software development courses. 
We tried to maximise diversity in terms of geographic regions/cultures as well as university settings.
We explicitly targeted rather experienced educators who had at least 5 years of teaching experience, since our goal was to collect feedback and reflect on the catalogue based on experience. 

The interviews were also conducted via zoom and lasted longer: between 30-40 minutes each, mainly because the introduction to the fun activities lasted longer (about 10 minutes) as the educators did not know them before (as opposed to the students). 
P10 and P11 were interviewed in the same session and worked at the same university. 
After a brief introduction and the consent collection, we asked the following open questions:

\begin{enumerate}
    \item Would you use (some of) the fun activities in your teaching and why?
    \item  What are, in your opinion, the main advantages and drawbacks of the fun activities?
    \item Do you have any specific feedback to one or more activities? Do you know similar activities?
\end{enumerate}
Similarly, we closed these interviews by asking if the educators have any other related comments or feedback to add.
We simultaneously typed the responses in a prepared form and followed up with clarifications and confirmed our notes at the end. 
We refrained from recording the interviews to ensure spontaneous, honest responses and protect the anonymity of the participants as agreed with them.

\subsection{Interview Results}
\subsubsection{Students' Perception and Stated Advantages}

\paragraph*{Overall perception}
Overall, all interviewed students appreciated the fun activities with nuances. 
Eleven of the 15 students expressed strong positive opinions about the fun activities, while the remaining four have  overall positive opinions with some reservations. 
Those four students said that some of the fun activities were  good and helpful for them, but not all.
Three students stated that they did not understand the NumNum reuse challenge at the time and felt it was somewhat unfitting. 
Two did not understand the paper planes activity. 
One student said he did not understand the rainstorm activity at the time but had an ``aha moment'' when we summarised it at the beginning of the interview session.
Several students highlighted that their favourite activities were the LA-OLA, linked lists, and the student sorting activities.  

When asking about the main advantages and how the activities helped, we noted the following recurring points:

\paragraph{Activation and focus}
Seven students mentioned that the fun activities helped them stay focused during the sessions. 
One student said: ``It helped me a lot, particularly to shake me up. 
Otherwise, it happens often in such courses that we lose the thread and our minds start going somewhere else.'' 
Another student said: ``Even if you try to stay focused, this is very hard for 90 minutes. The fun activities helped me to stay focused. I knew something was coming''. 
One student recommended conducting these fun activities in the second third of the lecture, where ``the focus is at the lowest level and people are most tired''.

\paragraph*{Understanding abstract concepts} 
Surprisingly, 14 out of the 15 students mentioned that the fun activities helped them to understand some key concepts in the course, either during the session of afterwards at home or while chatting with fellow students. 
One student said: ``When preparing for the exam, I think that the sorting game helped. I thought about the swaps. This was very good''. 
One other student stated: 
``I thought about the activities afterwards, what these examples aim at, and what the underlying concepts are. For instance, the infinite loop made me think. Also, the object-oriented thinking became much clearer''. 
One student claimed: ``Even if I did not understand 100\% at the moment what the goal of the activity was, it was helpful for the comprehension afterwards. For instance, I did not get the paper planes game at that moment. But, during the exercise session, I understood why we needed the cords and what this had to do with the pointers.''


Eight students highlighted that the fun activities particularly helped to \textit{visualise} the concepts. One said: ``I particularly found the visualisation helpful, even when not participating. This was good in addition to the visualisation on the slides and with the code examples.''
Three students also elaborated on the importance of varying the media and channels when learning foundational CS courses. One stated: ``It is important to me to get the same concept through different angles''. 
Another claimed: ``Particularly for the concepts that were new to  me [e.g. Linked List], this helped me to get started. Even for concepts I knew, it was helpful to see a different example and a different explanation method''. 
Two students also highlighted that they particularly appreciated the \textit{active participation} in the lecture to solve some tasks (in addition to the lab sessions) instead of passive content consumption. 

\paragraph*{Motivation and fun}
Eleven of fifteen students mentioned that a main advantage of the activities is that they kept them motivated, while nine students explicitly mentioned ``fun'' or ``nice atmosphere'' as a main advantage to them. 
One student said: ``Particularly at the beginning of the semester, it was very good to take away the tension and relax the course atmosphere. I went always with a very positive mood out of the lectures''.  

\paragraph*{Mnemonic aid}
Five students mentioned that some fun activities served as mnemonic aid and helped them memorise the concepts and the names.
One student said: ``I think the most important learning effect is that the activities presented a mnemonic aid to remember the concepts from the course, such as wiring, messages, pointers, or reuse''. 
One other student said: ``I could remember the concepts and their name much better than others. 
I particularly liked that some [key programming concepts] were highlighted with the fun activities''. 

\subsubsection{Students Perceived Drawbacks and Limitations}
\paragraph{Crystal clear instruction and goals}
Several participants mentioned that some of the fun activities were not completely clear to them: in terms of what is expected from them and the overall goal.  
Three students mentioned that they sometimes felt the activity wasn’t working perfectly, which they believe is crucial for its effectiveness. 
One student stated: ``For the paper plane, it was not fully clear to me what everyone should do. I think the instructions must be very precise to not confuse students even more''. 
Another student stated: ``Sometime, I think it did not work 100\%, such as the exit/termination condition with the LA-OLA wave. It might be because the instructions were not precise or because some students were not focused or did not understand the algorithm. 
To me, this was not an issue though. I got the learning point/metaphor and [the impreciseness] actually got me thinking about the details and how to improve it''.
One student stressed: ``The elements must work after two trials. Otherwise, they will be confusing. If it lasts longer (e.g. than five minutes) it will become boring. 
For instance, for traversing the tree in the hall, it might be better to show where exactly the tree is, e.g. on a slide''. 


Related to this point, five students highlighted that the explicit link to the remaining course material is key. 
One student said: ``There should always be a direct link to source code [examples] before or after''. 
Another student stressed the importance of always bridging  between the fun elements, the theory (i.e., slides), and practice (i.e., lab exercises, code examples, etc.). 
He continued: ``For instance, for the Linked List activity, I would have preferred to see the list implementation for this exact example.'' 
The students claimed that the explicit link and reflection are not only important to understand the fun examples but also to get convinced that the fun activities are ``actually important also for the exam and not just for fun'' as one stated.

\paragraph*{Fun activities require time}
Six participants commented that a potential drawback is the time needed to explain and run the activities, which would lack, e.g. for other examples and content. One student said he felt a few elements took too long. 
Another student explained that the time is particularly critical for those who know the content anyway like him (about half of the students had CS at schools in our course). 
Those might prefer to have advanced knowledge and details. 
One student said: ``One drawback might be the needed time. 
But for me, I would have done this even if it would have lasted longer''. 
One student highlighted that the activities might lead to a time issue, particularly if conducted at the end of the session at the cost of other content. 

\paragraph*{Inclined students}
Three students also mentioned that they felt or heard that some other students did not want to do the fun games and might have felt ``embarrassed''. 
One stated: ``Some do not want to participate or are not in the mood. You might loose them during the fun activity. Therefore, it's important to get them back afterwards, for instance, with the reflection or code example''. 
Another student said: ``Some might not want to do any activity. It is important to not force or embarrass them''. 

\paragraph*{Fun was not the main reason to attend in-person}
Our starting motivation for developing and conducting the fun activity was to make a difference and motivate students to attend physically despite the availability of high-quality video recordings. 
Therefore, we asked about this explicitly in the student interviews (Q4). 
However, almost all (except one) said that this was not the reason for them to attend regularly. 
Six students said that while this was not the main reason, the fun activities made their participation ``easier'' as they looked forward to attending the course. 
The main reasons to attend were to keep on track and the collective experience with others.
One student answered this question as follows: ``I think yes. I am not sure if it was the main reason. But it was great to get activated and for fun. I thought, it's cool to participate again''. 

\subsubsection{Educators Perception and Feedback}
Twelve out of the 14 interviewed educators stated that they would use the fun activities in their courses or at least part of them. 
P3, P5, P7, P8, P12, and P14 had used similar fun activities but with smaller classes. 
P3 and P13 mentioned the CS-Unplugged\footnote{\url{https://www.csunplugged.org/}} initiative \cite{Bell:18:Adventures}, which similarly collect fun games to teach computer science concepts, but rather for small groups and for school children than for university students. 
P5 tried earlier a similar activity like the binary calculator. P7 asked small groups of students to collectively resolve a logic task and made a competition out of it. P8 mentioned other design games like The Tower of Hanoi\footnote{\url{https://en.wikipedia.org/wiki/Tower_of_Hanoi}} with students representing the rings, Spaghetti Tower with marshmallows\footnote{\url{https://www.wikihow.com/Build-a-Spaghetti-Tower}} or using LEGO in their requirements engineering, design, and project management courses \cite{Gama:SBES:19}. 
Other educators also tried other activation elements like calling out students (P6), giving badges to correct answers (P5), having small discussion groups (P1), or bringing a real electrical adapter to the course when teaching the adapter pattern (P4). 

P1 stated: ``Particularly in the first 3-5 weeks, these activities are very good and [...] will work well to get the basics of programming. In the second half of the course when things get more complex, I would use different activation techniques, like building actual teams with small tasks, having them discuss with peers, hands-on examples, etc.''. 
He mentioned that the course he taught was very large but did not include data structure, had more software engineering elements, and the students were all CS majors, most of them with some previous experience. 
P3 mentioned that integrating the fun elements is ``way better than the boring way'' and highlighted particularly the fitness to classes with hundreds of students. 
He said: ``You should provide the catalogue. I would definitely use them''.

Only P6 had some reservations. He stated: ``the metaphors are clear and make sense; but I think there is a risk that students might backfire, particularly in a not-so-cooperative group.'' 
He highlighted previous experiences, where students complained boldly because they were called randomly and felt pointed out. 
He also highlighted that this likely depends on two main factors: (a) the expectations from the course, e.g.~what the students hear from their fellow students of previous years or read online, and (b) the culture/personality of the group.  
He continued: ``Even after years of teaching experience, I have some tension before teaching large classes every time. Because [...] good reputation is crucial for teaching success''. 
Therefore this educator is very careful with uncommon, likely misinterpretable teaching elements. 

Concerning the single elements, P1 and P2 gave the feedback that the NumNum reuse challenge is different as no physical activity was executed. 
P2 expected that the reuse example might be confusing with inheritance. 
P3, P7 and P8 asked how the paper planes (messages) were bound to the cords (references), which was not the case in our course, as the cords were set to establish the connections and the planes were sent ``roughly'' in the hall to hit the connected person. 
Apart from this, all elements were praised by the interviewees. 
P1 stated: ``I really love the [LA-OLA] wave. I will implement it in my course. It’s a perfect introduction to algorithms and control structures.   Message passing between objects: I love it too.''  
P2 said: ``The recursion game is spectacular as the students execute the whole algorithms by themselves and remember the execution parts [single recursions], multiple method activations, and the returns''.

\paragraph{Advantages as seen by educators}
Overall, the interviewed educators stated the same advantages and drawbacks as the students, while the details and frequencies/focus were slightly different. 
Nine of 15 educators (P3, P4, P5, P6, P8, P11, P12, P13, and P14) mentioned that the fun factor itself is a major advantage of the activities. 
Also, nine (P1, P2, P3, P6, P8, P9, P12, P13, and P14) discussed the increased motivation of students about the course and its content. 
P2 stated: ``In such introductory large lectures, a main goal is often to catch the people and fascinate them'' as this will motivate them to learn more and better and impact the remaining courses. P6 also noted: ``It’s hard to decouple learning from motivation''.

Interestingly, seven interviewed educators highlighted that one main advantage is that the activities help remembering the content (P1, P2, P5, P7, P8, P11, and P13). 
P1 said: ``I still remember the person that taught me programming because he gave good memorable examples about referencing. The biggest advantage is that students will remember the content [of your] course. They will also talk about it. Perhaps they become tutors or teachers themselves and they will consider doing it in the same or similar way.''
P7 stated: ``[Such activities] will engrave the information'' referring to the proverb ``Tell me and I forget, show me and I understand, involve me and I will remember''.
P5 stated: ``It’s hard to tell if it will really help with the understanding and learning. Some [students] if they are technical, might be annoyed. For others (first time to informatics) the activities might help with remembering and understanding''. 

Also, seven out of 15 highlighted the advantage of having different methods and  visualisations for a large, heterogeneous crowd. 
P4 stated: ``I once brought a real electricity adapter and fuses to explain architectural concepts. In other fields like physics or chemistry, it is quite common to see spectacular shows by educators to explain complex phenomena. 
In our field, this is rare and difficult, apart from live programming maybe.''
P2 put it this way: ``Having a different perspective/visualisation on the same concept is important, including different examples, teaching methods, or activation exercises. This not only addresses different learning styles and personalities but also supports and deepens the abstraction skills [of students]''.

\paragraph{Drawbacks and challenges seen by educators}
The drawbacks and challenges stated by the educators were also  similar to those stated by the interviewed students. 
All interviewees but P14 mentioned that some students might get lost or not willing to participate. 
P3 stated: ``Would students leave because they don’t like it? There are perhaps 1-2 in a couple of hundreds. They think it’s silly because they know everything. But that’s OK.''
P2 elaborated in detail about the importance of integrating such elements in the remaining course material, particularly to get those sceptical students or those who do not understand the abstractions and metaphors.

Time was also stated as a challenge by most interviewees but not by P6, P14. Interviewees repeatedly explained that the lectures are usually very dense and packed with content. 
P1 added: ``Attention spans of students are getting much smaller. We condensed lectures from 2x45 minutes to 1x45 minutes''. 
P7 and P8 also mentioned the effort for preparing and buying the stuff needed for conducting the fun activities. 

One interesting additional challenge that emerged from the educator interviews was \textit{potential cultural differences} or sensitivity (P6, P7, P8, and P9). 
P6 mentioned that in some cultures (also organisational or in groups) students might be very sceptical to every different method, particularly asking them to do things when they are usually not.  
P7 stated the opposite: ``Our students love anything apart from the slides, particularly if it's fun they will do whatever you ask for''. 

\subsection{Threats to Validity}
As in every empirical study, ours has limitations too. 
First, we used qualitative methods and aimed neither at quantification nor representativeness. 
Our goal was to rather to gather potential benefits and pitfalls for educators who want to use such activities.
Therefore, the observations from the interviews and from executing the fun activities three consecutive years might not generalise. Particularly, the ratios and frequencies are not meant for generalisability but rather to show the redundancies in independent statements of participants and to reveal trends. 

Only students from our foundational programming course participated in the interviews. 
Therefore, their opinions and feedback certainly do not generalise to other students from other universities or having other demographics. 
For the student interviews, it was necessary that the interviewees already have had experience and participated in the fun game to be able to assess whether and how it helped them or impacted their way of learning software development. 
The course style, culture, particular statements during the term, or experiences from the course might have impacted their overall assessments and opinions on the fun activities. 

We also recruited the lecturers from our network. 
Thus, they might be potentially biased, having similar teaching styles or preferences and thus representing a homogeneous sample of educators. 
We mitigated this potential threat by being as diverse as possible with the sample: targeting different universities around the world. 
We also focused on experienced educators. 
This enabled us to have in-depth reflections on the pros and cons and also reduced a potential confirmation bias of telling us what we would like to hear. 
Finally, we started our interviews by asking participants to be as  honest and as critical as possible. 
Nevertheless, conducting a broad survey potentially disseminated through CS education communities would certainly increase the generalisability of the results.

Moreover, volunteer and selection bias are common in subjective studies with volunteers and convenience sampling. 
To cope with such a limitation, we phrased our questions as neutral as possible, seeking for positive and negative aspects (which is well reflected in the results too). 
We also contacted the students only once the course and all exams terminated and ensured that their feedback remains anonymous and does not have any impact on any of their study activities. 
Similarly, interviewer bias is a common threat to validity, particularly as we refrained from recording the interview sessions. 
This was, however, crucial for authenticity as interviewees reflected partly on sensitive content, including discussing their own negative experience, describing undesired behaviour of own students or fellow students, etc. 
To limit this threat, we summarised our notes at the end of each interview section.

Finally, our evaluation focuses on the subjective opinions of selected students and educators. 
While this enables an in-depth understanding of the applicability of our fun activities, their advantages, and limitations in different contexts, a next step would be to design experiments and collect data to quantitatively check the observations and measure the effects, e.g. the ratios of inclined students or the optimal duration and timing of the activities. 

\section{Related Work}
\label{sec:relwork}

While fun is generally well-accepted as an effective pedagogical method in the education of children and seniors, it is rather rare in adult education literature, including university education \cite{Lucardie:PSBS:2014}. 
Lucardie argued earlier that this should be further investigated \cite{Lucardie:PSBS:2014}, which was one motivating cornerstone for developing our catalogue of fun software development activities with hundreds of students. 
In a study with college students, Tews et al.~\cite{Tews:CT:2015} investigated the impact of different fun activities such as small competitions, hands-on activities, field trips, playing music, or bringing food to lectures on student engagement. 
The authors found that specific activities (which the authors call fun delivery in particular creative examples, humour, storytelling, and attention getters) have a significant positive correlation with cognitive engagement.  The authors also found that these fun deliveries had a stronger impact on engagement than peer socialising or instructor praise. 
The fun activities described in Section \ref{sec:catalog} fall under the category of fun delivery and are likely to have a similar impact as our experience and interview study suggest. 
What is particular about our list of fun activities is that they are tailored to introducing and discussing basic programming and software development concepts. 
These are often abstract concepts, tightened to computers, and rather decoupled from real-life scenarios. 
Also, the physical, cost-effective scalability to hundreds of students is a key unique challenge to basic frontal courses. 


Over the last decade, games (both physical and digital) are becoming more and more common in software engineering education---particularly for smaller groups of students and particularly to reflect on software design and collaboration concepts. 
Rodríguez et al.~\cite{Rodriguez:CAEE:2021} recently conducted a literature review about using serious games for teaching Agile methods and identified 22 primary studies reporting on computer games, card games, or LEGO games. 
Gomes FM et al.~\cite{Gomes:SEET:17} reported on a case study of using game elements in a Software Engineering study group. 
Their students reported positive effects such as improved content comprehension, retention, and recap. The authors also reported similar challenges to ours, in particular difficulties related to a lack of time. 
Clegg et al.~\cite{Clegg:SEET:17} introduced the Code Defenders game,
to teach software testing concepts, particularly mutation testing, in a fun and competitive way. Their goal was to make software testing education more enjoyable.
Morales-Trujillo \cite{Morales-Trujillo:SEET:21} introduced KUALI-Brick, a  LEGO-based activity that bridges Software Quality Assurance and Software Process Improvement concepts, applying them in order to successfully build a LEGO city.
Similarly to our finding, the author highlighted the importance of  preparation as well as discussion at the end of the activity in order to emphasise the learning outcomes. 

De Almeida Souza et al.~\cite{DeAlmeidaSouza:SEET:17} identified 106 primary studies describing the use of serious games, gamification, and game development in software engineering education. 
By categorising the papers, the authors found that ``Software Process'' and ``Software Design'' are the most recurring knowledge areas explored by game-related approaches, while areas directly related to source code and programming (software verification, quality, and analysis) were the least. 
A similar gap was observed by Miljanovic et al.~\cite{Miljanovic:SG:18}. The authors surveyed 49 of such games and concluded that there is a lack of games that focus on data structures, development methods, and software design. 
Our catalogue reduces this gap.

There is also a large body of knowledge around the area of interactive learning and student activation in large classrooms. 
Krusche et al.~\cite{Krusche:ACEC:2017} suggested interactive learning to increase student participation in software programming courses through shorter exercise cycles on computers. 
Kothiyal et al. \cite{Kothiyal:ICER:13}  think-pair-share a classroom-based active learning strategy, in which students work on a problem posed by the instructor, first individually, then in pairs, and finally as a class-wide discussion. The authors studied the effect of this learning strategy in a large CS1 class and concluded a large ratio of sustained engagement.
Krusche et al.~\cite{Krusche:SEET:17} introduced a self-organising, adaptive, and non-linear learning approach called  Chaordic Learning to stimulate the creative thinking of CS students. Applied in a games development course, they observed that the approach increased intrinsic motivation, an improved learning experience.
Knobloch et al.~\cite{Knobloch:SEET:18} introduced a teaching framework to increase interaction of CS students by parallelising it with content delivery and  lowering barriers for students to participate. 
The authors showed an increase in student participation when applying their framework 
leading to increased student examination performance for active students.
All these approaches are complementary to our fun catalogue as they address different student personalities and barely focus on fun and easy-memorable experiences. 

Student response systems (SRS) represent perhaps one of the most popular and best studied activation and interaction techniques in larger classrooms. 
Denker \cite{Denker:CT:13} explored the effectiveness of incorporating student response systems into large lectures. 
The author found that SRS encourages greater participation. 
Students' perceptions of empowerment and motivation in the classroom impact their views of clickers, given that the more empowered/motivated students felt that they were both more engaged and learning more. 
This is consistent with our findings as we found that fun activities will likely increase the student motivation and empowerment/interaction perception.
Dolezal et al.~\cite{Dolezal:EdMedia:18} 
studied the impact of the game-based student response systems on flipped introduction to programming courses at universities. The authors observed  a significant improvement in the satisfaction with the course structure, course design, and perceived suitability of the course material.
Cicirello \cite{Cicirello:CSE:09} studied the role and effectiveness of unannounced ``pop'' quizzes in CS1. 
The author found that CS and Math majors both receive a greater benefit from pop quizzes than other non-majors.
This highlights the relevance for other activation exercises and student engagement techniques, particularly for mixed CS1 courses (as it was the case in our course).

\section{Discussion and Conclusion}
\label{sec:discusion}
We presented and discussed a catalogue of ten novel fun activities for visualising, memorising, and reflecting upon basic software development concepts. The fun activities include: 
\begin{itemize}
    \item ``Executing'' algorithms, control structures, and Boolean expressions with hundreds of students.
    \item ``Playing'' fundamental object-oriented concepts such as object states and behaviours, object references and networks, and software reuse. 
    \item Simulating and visualising data structures and algorithms, e.g., with claps and finger pointing. 
\end{itemize}  

Our catalogue particularly targets large, frontal, plenary sessions, where other interaction methods such as Q\&A or group work might be difficult to conduct. 
It is not meant to be complete and can be extended or tuned depending on the size of the class, the teaching content, or the available time and resources. 
The activities as they are presented in this paper are developed over years and are specifically tailored to software development concepts. 
Variations of some activities were originally observed in large  performance halls, without a link to programming and software development. 
We developed the appropriate metaphors to the teaching content together with the supporting materials (like code examples). 
Other activities were common small games (e.g.~paper planes or calculators) which we extended and adjusted to scale up as presented in the paper. 
By sharing our catalogue, we hope that others will use and extend the elements, hoping to build a community around it.   
This can help overcome the difficulties in isolating the effect of single activities on learning performance and increase the coverage to other topics such as inheritance, architectural patterns, testing, and Agile. 
In general, while our interview studies revealed many insights about the pros and cons and as well as potential impacts and limitations of the fun activities, a systematic quantitative evaluation over multiple courses, universities, and years would be needed to better understand the impact of these fun activities on the learning progress and outcome.

Our interview study suggests that professors should carefully consider the fitness of the activities to the learning goal and the classroom setting. 
They might find some of the activities inappropriate for the course content or ``culture''.  
Moreover, as lack of time and dense content is one of the main challenges of such classrooms, the activities should remain short, e.g. lasting not longer than 8-12 minutes during a 90-minute session. 
This would roughly correspond to a short break. 
Professors should also be prepared that the activities might take more time than planned as unexpected questions or events might arise. 
Ideally, the activities should be conducted in the middle of the session, in order to physically activate the student and relax the setting. 
If conducted at the beginning, they might take longer and skew the whole session. 
If conducted at the end, the activation effect gets lost, and some students might simply leave earlier to avoid them (e.g. when asking for volunteers). 

Students might sometimes come across as passive and refrain from engaging and participating. 
For the successful deployment of these activities, it is crucial to establish an open and fun culture from the beginning of the course (but serious and respectful). 
The LA-OLA activity is a good ice breaker which sets the expectation from week 1. 
Also, for an effective execution of the fun activities, it is crucial to share with the students clear and precise instructions, ideally on a slide showing the expected outcome. 
Otherwise, the activity might lead to misunderstanding, multiple clarification questions, and chaos in the worst case. 

Finally, we think that student engagement methods, including but not limited to fun activities, might make the main difference for in-presence classrooms in the future.
Obviously, fun activities represent \textit{only one} of many pedagogical methods for activating and engaging students \cite{Tews:CT:2015}. 
Our catalogue is likely \textit{not enough} to achieve a high learning outcome. 
Other methods include: 
\begin{itemize}
    \item Storytelling from the lecturer’s experience or from famous cases that show the importance of software engineering~\cite{Martin:22}. 
    \item Programming demos showing, e.g.~concrete code examples in the IDE, in the debugger, or e.g. in Karel the Robot \cite{Pattis:1994}.
    \item Online questions, e.g.~in Mentimeter to regularly collect feedback about the overall student progress and difficult topics, as well as multiple-choice questions to repeat content. 
    \item Bringing small cookies or items of appreciation in certain milestones or festivities (e.g. apples or oranges last session before the holiday season). 
\end{itemize}   
The case course discussed in this paper applied all those methods.   
In addition to the weekly plenary lectures, the students also participated in a weekly programming lab, where they work in pairs on small programming assignments to practice theory. 

\begin{acks}
We would like to thank all SE1 students at the University of Hamburg who participated in the games for their open-mindedness. We are also grateful all interviewees for their time and feedback. 
A variant of the LA-OLA activity was first observed in a large course by Bernd Bruegge, whose engaging and fun teaching style has significantly inspired the author and this work in general.

\end{acks}

\balance
\bibliographystyle{ACM-Reference-Format}
\bibliography{ref}

@article{Maddox:NMS:2021,
author = {Jessica Maddox},
title ={The secret life of pet Instagram accounts: Joy, resistance, and commodification in the Internet’s cute economy},
journal = {New Media \& Society},
volume = {23},
number = {11},
pages = {3332-3348},
year = {2021},
doi = {10.1177/1461444820956345}
}

@article{Gleason:CT:1986,
  title={Better communication in large courses},
  author={Gleason, Maryellen},
  journal={College Teaching},
  volume={34},
  number={1},
  pages={20--24},
  year={1986},
  publisher={Taylor \& Francis}
}

@article{Lucardie:PSBS:2014,
  title={The impact of fun and enjoyment on adult's learning},
  author={Lucardie, Dorothy},
  journal={Procedia-Social and behavioral sciences},
  volume={142},
  pages={439--446},
  year={2014},
  publisher={Elsevier}
}

@article{Tews:CT:2015,
  title={Fun in the college classroom: Examining its nature and relationship with student engagement},
  author={Tews, Michael J and Jackson, Kathy and Ramsay, Crystal and Michel, John W},
  journal={College Teaching},
  volume={63},
  number={1},
  pages={16--26},
  year={2015},
  publisher={Taylor \& Francis}
}

@inproceedings{Krusche:ACEC:2017,
  title={Interactive learning: Increasing student participation through shorter exercise cycles},
  author={Krusche, Stephan and Seitz, Andreas and B{\"o}rstler, J{\"u}rgen and Bruegge, Bernd},
  booktitle={Proceedings of the Nineteenth Australasian Computing Education Conference},
  pages={17--26},
  year={2017}
}

@book{Pattis:1994,
  title={Karel the robot: a gentle introduction to the art of programming},
  author={Pattis, Richard E},
  year={1994},
  publisher={John Wiley \& Sons}
}

@article{Kay:CE:2009,
  title={Examining the benefits and challenges of using audience response systems: A review of the literature},
  author={Kay, Robin H and LeSage, Ann},
  journal={Computers \& Education},
  volume={53},
  number={3},
  pages={819--827},
  year={2009},
  publisher={Elsevier}
}

@article{Mayhew:RLT:2020,
  title={The impact of audience response platform Mentimeter on the student and staff learning experience},
  author={Mayhew, Emma and Davies, Madeleine and Millmore, Amanda and Thompson, Lindsey and Pena, Alicia},
  journal={Research in Learning Technology},
  volume={28},
  year={2020},
  publisher={Association for Learning Technology}
}

@inproceedings{Zhang:ITCSE:2022,
  title={A Case Study of Middle Schoolers' Use of Computational Thinking Concepts and Practices during Coded Music Composition},
  author={Zhang, Yifan and Krug, Douglas Lusa and Mouza, Chrystalla and Shepherd, David C and Pollock, Lori},
  booktitle={Proceedings of the 27th ACM Conference on on Innovation and Technology in Computer Science Education Vol. 1},
  pages={33--39},
  year={2022}
}

@article{Rodriguez:CAEE:2021,
  title={Serious games for teaching agile methods: A review of multivocal literature},
  author={Rodr{\'\i}guez, Guillermo and Gonz{\'a}lez-Caino, Pablo C and Resett, Santiago},
  journal={Computer Applications in Engineering Education},
  volume={29},
  number={6},
  pages={1931--1949},
  year={2021},
  publisher={Wiley Online Library}
}

@article{Zhu:OL:2018,
  title={Instructor experiences designing MOOCs in higher education: Pedagogical, resource, and logistical considerations and challenges.},
  author={Zhu, Meina and Bonk, Curtis J and Sari, Annisa R},
  journal={Online Learning},
  volume={22},
  number={4},
  pages={203--241},
  year={2018},
  publisher={ERIC}
}

@misc{MyPath:22,
    author = {MyPath Education}, 
    title =  {6 Benefits of Face-To-Face Learning}, 
    year = 2022,
    month = 4,
    url = {https://mypatheducation.edu.au/health-and-aged-care-courses/6-benefits-of-face-to-face-learning/}
}

@misc{Martin:22,
    author = {Daniel Martin}, 
    title =  {11 of the most costly software errors in history}, 
    year = 2022,
    month = 4,
    publisher = {Raygun Blog Post},
    url = {https://raygun.com/blog/costly-software-errors-history/}
}

@article{Rocca:CE:2010,
  title={Student participation in the college classroom: An extended multidisciplinary literature review},
  author={Rocca, Kelly A},
  journal={Communication education},
  volume={59},
  number={2},
  pages={185--213},
  year={2010},
  publisher={Taylor \& Francis}
}

@article{Sass:TP:1989,
  title={Motivation in the college classroom: What students tell us},
  author={Sass, Edmund J},
  journal={Teaching of psychology},
  volume={16},
  number={2},
  pages={86--88},
  year={1989},
  publisher={SAGE Publications Sage CA: Los Angeles, CA}
}

@article{Douglas:AER:2012,
  title={Digital devices, distraction, and student performance: Does in-class cell phone use reduce learning},
  author={Duncan, Douglas K and Hoekstra, Angel R and Wilcox, Bethany R},
  journal={Astronomy education review},
  volume={11},
  number={1},
  pages={1--4},
  year={2012}
}

@misc{Leustig:19,
    author = {Tierra Leustig}, 
    title =  {6 of the Most Common Classroom Distractions}, 
    year = 2019,
    month = 12,
    url = {https://www.dyknow.com/blog/6-of-the-most-common-classroom-distractions/}
}

@inproceedings{Zhang:ICSE:2018,
  title={Are code examples on an online q\&a forum reliable? a study of api misuse on stack overflow},
  author={Zhang, Tianyi and Upadhyaya, Ganesha and Reinhardt, Anastasia and Rajan, Hridesh and Kim, Miryung},
  booktitle={Proceedings of the 40th international conference on software engineering},
  pages={886--896},
  year={2018}
}

@inproceedings{Haberman:ITCSE:02,
  title={The case of base cases: Why are they so difficult to recognize? Student difficulties with recursion},
  author={Haberman, Bruria and Averbuch, Haim},
  booktitle={Proceedings of the 7th annual conference on innovation and technology in computer science education},
  pages={84--88},
  year={2002}
}

@article{Lacave:TE:17,
  title={A preliminary instrument for measuring students’ subjective perceptions of difficulties in learning recursion},
  author={Lacave, Carmen and Molina, Ana I and Redondo, Miguel A},
  journal={IEEE Transactions on Education},
  volume={61},
  number={2},
  pages={119--126},
  year={2017},
  publisher={IEEE}
}

@inproceedings{Shaffer:CSE:07,
  title={Algorithm visualization: a report on the state of the field},
  author={Shaffer, Clifford A and Cooper, Matthew and Edwards, Stephen H},
  booktitle={Proceedings of the 38th SIGCSE technical symposium on Computer science education},
  pages={150--154},
  year={2007}
}

@article{Bell:18:Adventures,
  title={CS unplugged—how is it used, and does it work?},
  author={Bell, Tim and Vahrenhold, Jan},
  journal={Adventures between lower bounds and higher altitudes: essays dedicated to Juraj Hromkovi{\v{c}} on the occasion of his 60th birthday},
  pages={497--521},
  year={2018},
  publisher={Springer}
}

@inproceedings{Gama:SBES:19, author = {Gama, Kiev}, title = {An Experience Report on Using LEGO-Based Activities in a Software Engineering Course}, year = {2019}, isbn = {9781450376518}, publisher = {Association for Computing Machinery}, address = {New York, NY, USA}, url = {https://doi.org/10.1145/3350768.3353817}, doi = {10.1145/3350768.3353817}, booktitle = {Proceedings of the XXXIII Brazilian Symposium on Software Engineering}, pages = {289–298}, numpages = {10}, location = {Salvador, Brazil}, series = {SBES '19} }

@inproceedings{Kothiyal:ICER:13, 
author = {Kothiyal, Aditi and Majumdar, Rwitajit and Murthy, Sahana and Iyer, Sridhar}, 
title = {Effect of Think-Pair-Share in a Large CS1 Class: 83\% Sustained Engagement}, 
year = {2013}, isbn = {9781450322430}, publisher = {Association for Computing Machinery}, 
address = {New York, NY, USA}, url = {https://doi.org/10.1145/2493394.2493408}, 
doi = {10.1145/2493394.2493408}, 
booktitle = {Proceedings of the Ninth Annual International ACM Conference on International Computing Education Research}, pages = {137–144}, numpages = {8}, location = {San Diego, San California, USA}, series = {ICER '13} 
}

@inproceedings{Gomes:SEET:17,
  author={Gomes Fernandes Matsubara, Patrícia and Lima Corrêa Da Silva, Caroline},
  booktitle={2017 IEEE/ACM 39th International Conference on Software Engineering: Software Engineering Education and Training Track (ICSE-SEET)}, 
  title={Game Elements in a Software Engineering Study Group: A Case Study}, 
  year={2017},
  volume={},
  number={},
  pages={160-169},
  doi={10.1109/ICSE-SEET.2017.8}
}

@INPROCEEDINGS{Krusche:SEET:17,
  author={Krusche, Stephan and Bruegge, Bernd and Camilleri, Irina and Krinkin, Kirill and Seitz, Andreas and Wöbker, Cecil},
  booktitle={2017 IEEE/ACM 39th International Conference on Software Engineering: Software Engineering Education and Training Track (ICSE-SEET)}, 
  title={Chaordic Learning: A Case Study}, 
  year={2017},
  volume={},
  number={},
  pages={87-96},
  doi={10.1109/ICSE-SEET.2017.21}}

@INPROCEEDINGS{Clegg:SEET:17,
  author={Clegg, Benjamin S. and Rojas, Jose Miguel and Fraser, Gordon},
  booktitle={2017 IEEE/ACM 39th International Conference on Software Engineering: Software Engineering Education and Training Track (ICSE-SEET)}, 
  title={Teaching Software Testing Concepts Using a Mutation Testing Game}, 
  year={2017},
  volume={},
  number={},
  pages={33-36},
  doi={10.1109/ICSE-SEET.2017.1}}

@INPROCEEDINGS{DeAlmeidaSouza:SEET:17,
  author={De Almeida Souza, Mauricio Ronny and Furtini Veado, Lucas and Teles Moreira, Renata and Magno Lages Figueiredo, Eduardo and Costa, Heitor Augustus Xavier},
  booktitle={2017 IEEE/ACM 39th International Conference on Software Engineering: Software Engineering Education and Training Track (ICSE-SEET)}, 
  title={Games for learning: bridging game-related education methods to software engineering knowledge areas}, 
  year={2017},
  volume={},
  number={},
  pages={170-179},
  doi={10.1109/ICSE-SEET.2017.17}}

@InProceedings{Miljanovic:SG:18,
author="Miljanovic, Michael A.
and Bradbury, Jeremy S.",
editor="G{\"o}bel, Stefan
and Garcia-Agundez, Augusto
and Tregel, Thomas
and Ma, Minhua
and Baalsrud Hauge, Jannicke
and Oliveira, Manuel
and Marsh, Tim
and Caserman, Polona",
title="A Review of Serious Games for Programming",
booktitle="Serious Games",
year="2018",
publisher="Springer International Publishing",
address="Cham",
pages="204--216",
isbn="978-3-030-02762-9"
}

@INPROCEEDINGS{Morales-Trujillo:SEET:21,
  author={Morales-Trujillo, Miguel Ehécatl},
  booktitle={2021 IEEE/ACM 43rd International Conference on Software Engineering: Software Engineering Education and Training (ICSE-SEET)}, 
  title={Learning Software Quality Assurance with Bricks}, 
  year={2021},
  volume={},
  number={},
  pages={11-19},
  doi={10.1109/ICSE-SEET52601.2021.00010}}

@inproceedings{Knobloch:SEET:18, author = {Knobloch, Jan and Kaltenbach, Jonas and Bruegge, Bernd}, title = {Increasing Student Engagement in Higher Education Using a Context-Aware Q\&A Teaching Framework}, year = {2018}, isbn = {9781450356602}, publisher = {Association for Computing Machinery}, address = {New York, NY, USA}, url = {https://doi.org/10.1145/3183377.3183389}, doi = {10.1145/3183377.3183389}, booktitle = {Proceedings of the 40th International Conference on Software Engineering: Software Engineering Education and Training}, pages = {136–145}, numpages = {10}, location = {Gothenburg, Sweden}, series = {ICSE-SEET '18} }

@inproceedings{Dolezal:EdMedia:18,
          author = {D. Dolezal and A. Posekany and Renate Motschnig and T. Kirchweger and R. Pucher},
           month = {June},
            year = {2018},
       booktitle = {EdMedia + Innovative Learning Conference 2018, Amsterdam, Netherlands, 25-29 June 2018},
           title = {Impact of Game-Based Student Response Systems on Factors of Learning in a Person-Centered Flipped Classroom on C Programming},
             url = {http://eprints.cs.univie.ac.at/5963/}
}

@article{Denker:CT:13,
author = {Katherine J. Denker},
title = {Student Response Systems and Facilitating the Large Lecture Basic Communication Course: Assessing Engagement and Learning},
journal = {Communication Teacher},
volume = {27},
number = {1},
pages = {50-69},
year = {2013},
publisher = {Routledge},
doi = {10.1080/17404622.2012.730622}
}

@inproceedings{Cicirello:CSE:09, author = {Cicirello, Vincent A.}, title = {On the Role and Effectiveness of Pop Quizzes in CS1}, year = {2009}, isbn = {9781605581835}, publisher = {Association for Computing Machinery}, address = {New York, NY, USA}, url = {https://doi.org/10.1145/1508865.1508971}, doi = {10.1145/1508865.1508971}, booktitle = {Proceedings of the 40th ACM Technical Symposium on Computer Science Education}, pages = {286–290}, numpages = {5}, location = {Chattanooga, TN, USA}, series = {SIGCSE '09} }

@inproceedings{Pham:ICSEW:18, 
author = {Pham, Yen Dieu and Fucci, Davide and Maalej, Walid}, 
title = {A first implementation of a design thinking workshop during a mobile app development course project}, 
year = {2018}, 
isbn = {9781450357500}, 
publisher = {Association for Computing Machinery}, address = {New York, NY, USA}, 
url = {https://doi.org/10.1145/3194779.3194785}, 
doi = {10.1145/3194779.3194785}, 
booktitle = {Proceedings of the 2nd International Workshop on Software Engineering Education for Millennials}, pages = {56–63}, numpages = {8}, location = {Gothenburg, Sweden}, series = {SEEM '18} }

\end{document}